\documentclass[11pt]{article}
\usepackage{amsmath,amsthm,amssymb,amscd}
\usepackage{hyperref}
\usepackage{geometry}
\usepackage{tikz-cd}
\usepackage{booktabs} 
\usepackage{graphicx} 
\usepackage{multirow}
\usepackage{placeins}
\usepackage{tikz}
\usepackage{xcolor}
\usepackage{authblk}
\usetikzlibrary{calc,positioning,3d}
\usepackage[normalem]{ulem} 
\usepackage{algorithm}
\usepackage[font=footnotesize]{caption}
\usepackage{subcaption}
\usepackage{algpseudocode}%
\algrenewcommand\textproc{}
\usepackage{imakeidx}
\usepackage{authblk}

\makeindex

\usepackage{xcolor}

\newcommand{\keywords}[1]{%
    \textbf{Keywords:} #1
}

\title{OTMol: Robust Molecular Structure Comparison via Optimal Transport}

\author[1,$\dagger$]{Xiaoqi Wei}
\author[3,$\dagger$]{Xuhang Dai}
\author[5]{Yaqi Wu}
\author[5]{Yanxiang Zhao}
\author[3,4,*]{Yingkai Zhang}
\author[1,2,*]{Zixuan~Cang}

\affil[1]{Department of Mathematics, North Carolina State University}
\affil[2]{Center for Research in Scientific Computation, North Carolina State University}
\affil[3]{Department of Chemistry, New York University}
\affil[4]{Simons Center for Computational Physical Chemistry, New York University}
\affil[5]{Department of Mathematics, The George Washington University}

\date{}

\begin{document}

\footnotetext[2]{These authors contributed equally.}
\footnotetext[1]{Correspondence: zcang@ncsu.edu, yingkai.zhang@nyu.edu.}

\maketitle
\begin{abstract}
Root-mean-square deviation (RMSD) is widely used to assess structural similarity in systems ranging from flexible ligand conformers to complex molecular cluster configurations. Despite its wide utility, RMSD calculation is often challenged by inconsistent atom ordering, indistinguishable configurations in molecular clusters, and potential chirality inversion during alignment. These issues highlight the necessity of accurate atom-to-atom correspondence as a prerequisite for meaningful alignment. Traditional approaches often rely on heuristic cost matrices combined with the Hungarian algorithm, yet these methods underutilize the rich intra-molecular structural information and may fail to generalize across chemically diverse systems. In this work, we introduce OTMol, a method that formulates the molecular alignment task as a fused supervised Gromov-Wasserstein (fsGW) optimal transport problem. By leveraging the intrinsic geometric and topological relationships within each molecule, OTMol eliminates the need for manually defined cost functions and enables a principled, data-driven matching strategy. Importantly, OTMol preserves key chemical features such as molecular chirality and bond connectivity consistency. We evaluate OTMol across a wide range of molecular systems, including Adenosine triphosphate, Imatinib, lipids, small peptides, and water clusters, and demonstrate that it consistently achieves low RMSD values while preserving computational efficiency. Importantly, OTMol maintains molecular integrity by enforcing one-to-one mappings between entire molecules, thereby avoiding erroneous many-to-one alignments that often arise in comparing molecular clusters. Our results underscore the utility of optimal transport theory for molecular alignment and offer a generalizable framework applicable to structural comparison tasks in cheminformatics, molecular modeling, and related disciplines.

\noindent
\end{abstract}

\section{Introduction}

The spatial arrangement of atoms in a molecular system determines its physical properties, chemical reactivity, and biological activity \cite{johnson1990concepts, Yamamoto_2025}. With the continued advancement of experimental techniques such as X-ray crystallography, NMR spectroscopy, and cryo-electron microscopy, high-resolution molecular structures are now being determined at scale. Simultaneously, the growth of machine learning–based generative models \cite{AF3_Jumper2025_AlphaFold3,NeuPL_Qiao2024_NeuralPlexer, Boltz1_Wohlwend2024_Boltz1,Boltz2_Wohlwend2025_Boltz2, AF3_Xing2025_StateAware_AF3}, cheminformatics tools like RDKit, and molecular dynamics simulations has led to an exponential expansion of in silico 3D molecular datasets. Consequently, there is a growing need to accurately compare and align molecular structures.

Root-mean-square deviation (RMSD) is the most widely used metric for quantifying structural similarity. It plays a central role in conformer clustering, evaluating computational models, and identifying conserved structural motifs in molecular clusters. RMSD is also particularly useful for detecting redundancy in conformational ensembles \cite{RMSD_Lu_2019, RMSD_Lu_2021}, and for benchmarking predicted structures against experimental references.

Despite its broad application, RMSD-based comparison often faces practical challenges. A major issue is inconsistent atom ordering \cite{Yamamoto_2025}, which can occur when structures are generated by different software tools or experimental pipelines. Another challenge arises from symmetric or indistinguishable configurations, especially in molecular clusters such as water networks \cite{watercluster_hields2010, watercluster_Temelso2011} or noble gas assemblies \cite{noblegas_WalesDoye1997, noblegas_WalesCCD,noblegas_Xiang2004b, noblegas_Xiang2004a}, where multiple equivalent atom mappings exist. Additionally, some alignment algorithms allow reflections \cite{temelso2017arbalign}, which can invert molecular chirality, an outcome that is chemically and biologically undesirable. These challenges underscore the need for molecular structure comparison algorithms that are atom-order-independent, cluster-aware, and chirality-sensitive.

Alignment-based RMSD calculation requires identifying an optimal rigid-body transformation that minimizes atomic coordinate differences. Specifically, the superimposition of two molecules involves a rigid body transformation on the target molecule and aligning it to the reference molecule. When a one-to-one assignment of the atoms between the molecules is given, the optimal superimposition that has the lowest RMSD can be determined by the Kabsch algorithm \cite{kabsch1976solution} or the faster quaternions \cite{coutsias2004using}.
Since different atom assignments often yield substantially different superimpositions and RMSD values, it is crucial to identify a proper one-to-one atom assignment. To ensure that the RMSD accurately reflects the similarity between the two molecules, researchers seek the assignment that yields the lowest possible RMSD after optimal alignment. In the case of two molecules with the same atomic composition, the atom assignment problem is equivalent to the problem of finding permutations of the target molecule.
As the number of possible permutations of atoms is astronomical when the total number of atoms increases, naive approaches that brute-forcely evaluate all possible permutations are infeasible.

Many existing approaches \cite{allen2014implementation, helmich2012similarity, marques2010different, richmond2004alignment, sadeghi2013metrics, vasquez2009discovery, wagner2017armsd} find the atom assignment by solving a linear assignment problem \cite{burkard1999linear}. These approaches first construct a cost matrix that quantifies the dissimilarity between atoms of the two molecules. Then, the Hungarian algorithm \cite{kuhn1955hungarian, munkres1957algorithms} or the Jonker-Volgenant algorithm \cite{jonker1988shortest} is used to find a one-to-one atom assignment that minimizes the total assignment cost. A key challenge of solving a linear assignment problem in this context is the lack of a known cost matrix \textit{a priori}. Existing approaches address this by constructing heuristic cost matrices based on atomic properties, which often lead to suboptimal alignment and inflated RMSD values. For example, ArbAlign \cite{temelso2017arbalign} first performs coordinate transforms based on an intrinsic coordinate axis system and obtains an initial coarse superimposition. The cost of matching atoms is defined by their Euclidean distance. ArbAlign also applies coordinate swaps and reflections to generate additional initial superimpositions. An alternative coordinate transformation is proposed by Helmich and Sierka \cite{helmich2012similarity}. V{\'a}squez-P{\'e}rez \textit{et al.} \cite{vasquez2009discovery} and Marques \textit{et al.} \cite{marques2010different} apply the Hungarian algorithm with randomly generated rigid body transformations and iteratively improve the RMSD. Inspired by shape matching approaches in computer vision, Richmond \textit{et al.} \cite{richmond2004alignment} first define a histogram for each atom that encodes local spatial information, and then calculate the cost of matching two atoms as a weighted sum of the difference of partial charges and the difference of the histograms.

The assignment problem solved by the Hungarian algorithm can be viewed as a special case of optimal transport where the transport plan is restricted to permutation matrices. Here, we cast the problem of superimposing two molecules into a more general optimal transport problem. Conventional optimal transport in the Kantorovich form seeks a transport matrix which is also a probability matrix that minimizes the total transport cost based on a predefined inter-distribution cost. An important recent extension of optimal transport is the Gromov-Wasserstein optimal transport \cite{memoli2011gromov, peyre2016gromov}. Unlike traditional optimal transport, the Gromov-Wasserstein problem aims to preserve the intra-distribution distances. Intuitively, the Gromov-Wasserstein problem matches spatially close (or spatially distant, respectively) atoms in molecule A to spatially close (or spatially distant, respectively) atoms in molecule B. This approach eliminates the need to define a cost matrix \textit{a priori}. In this work, we apply Gromov–Wasserstein optimal transport to capture intra-molecular structural relationships, in contrast to existing methods that focus solely on inter-molecular distances.

Fundamental chemical principles impose inherent constraints on how molecules can be aligned. In practice, atom assignments often require certain atom categories or substructures to be matched. These chemistry constraints can be naturally incorporated into optimal transport formulations using the framework of supervised optimal transport (sOT) \cite{cang2022supervised} and supervised Gromov-Wasserstein optimal transport (sGW) \cite{cang2024supervised}. Specifically, the sOT framework allows elementwise constraints on the transport matrix. The sGW framework enforces structural consistency by constraining how much pairwise distances within each structure can be distorted during transport. Based on these recent optimal transport methods, we introduce OTMol, a computational tool for solving the molecular alignment problem. To ensure that atom labels match and utilize the intra-molecule structural information, we consider the fused supervised Gromov-Wasserstein (fsGW) problem, which combines sOT and sGW, addressing the atomic label costs and the molecular structure preservation, respectively. Moreover, OTMol can solve the alignment problem between clusters of atoms or molecules. In clusters of identical molecules, OTMol performs a hierarchical alignment by first assigning molecule pairs and then matching atoms within each pair, ensuring that atomic correspondences are restricted to their designated molecular matches.

The remainder of the paper is organized as follows. In section \ref{methods}, we first formalize the alignment problem and introduce the fused supervised Gromov–Wasserstein framework, which forms the mathematical foundation of OTMol. We then describe the OTMol algorithms for aligning individual molecules or clusters of identical atoms or molecules. Crucially, this framework is designed to overcome the core challenges identified above by ensuring that molecular alignment is atom-order-independent, cluster-aware, and chirality-sensitive, three essential properties for chemically meaningful structure comparison. In section \ref{results}, we compare the results of OTMol and ArbAlign, the state-of-the-art method for aligning molecules, on various benchmark datasets, including ATP, Imatinib, peptides, lipids, sugars, and water clusters. In \ref{discussion}, we outline future research directions on generalization to molecular analogs and macro-biomolecules.

\section{Materials and Methods}\label{methods}

In this section, we present the optimal transport problems related to OTMol. More details on general optimal transport background can be found in \cite{cang2022supervised, cang2024supervised, peyre2019computational, titouan2019optimal}.

\subsection{Mathematical Statement of the Molecular Alignment Problem}

Let A, B be two molecules of $n$ atoms and their coordinates be $\mathbf{x}_i^{\mathrm{A}}, \mathbf{x}_i^{\mathrm{B}}$. An atom assignment can be represented by a permutation matrix $\mathbf{P}$. We consider the following optimization problem
\begin{align}\label{eq:original_problem}
    \min_{\mathbf P} \min_{\mathbf{T}}\left(\frac{1}{n}\sum_{i,j} \mathbf{P}_{ij}\| \mathbf{x}^{\mathrm{A}}_i-\mathbf{T}(\mathbf{x}^{\mathrm{B}}_j)\|_2^2\right)^{\frac{1}{2}},
\end{align}where $\mathbf{T}$ is a rigid body transformation (reflection not allowed) in the 3-dimensional space. Given a permutation matrix $\mathbf{P}$, the Kabsch algorithm \cite{kabsch1976solution, lawrence2019purely, umeyama1991least} finds a rigid body transformation $\mathbf{T}$ such that the RMSD 
\begin{align}
    \left(\frac{1}{n}\sum_{i,j} \mathbf P_{ij}\| \mathbf{x}^{\mathrm{A}}_i-\mathbf{T}(\mathbf{x}^{\mathrm{B}}_j)\|_2^2\right)^{\frac{1}{2}}
\end{align}is minimized. 
Since the number of possible $\mathbf{P}$ is $n!$, the challenge is to find an approximate best permutation $\mathbf{P}$ in polynomial time. In addition, the permutation should also satisfy chemistry constraints.

\subsection{Fused Supervised Gromov-Wasserstein Optimal Transport}

Here we aim to approximate the optimal permutation by solving a Gromov-Wasserstein (GW) problem which seeks continuous transport plans preserving the pairwise structural relationships:
\begin{align}\label{problem:gw}
    \mathop{\arg \min}_{\mathbf P\in \Gamma} \sum_{i,j,k,l} (\mathbf D^{\mathrm{A}}_{ik} - \mathbf D^{\mathrm{B}}_{jl})^2\mathbf P_{ij}\mathbf P_{kl}
\end{align}where $\Gamma = \{\mathbf P \in \mathbb R^{n \times n}_+: \mathbf P\mathbf 1_n = \frac{1}{n}\mathbf 1_n, \mathbf P^T\mathbf{1}_n = \frac{1}{n} \mathbf{1}_n\}$ ($\mathbf 1_n$ is a vector of ones). $\mathbf P_{ij}$ can be interpreted as the probability of assigning the $i$-th atom in A to the $j$-th atom in B. The distance matrices $\mathbf D^\mathrm{A}$ and $\mathbf D^\mathrm{B}$ encode the intra-molecular geometrical and structural information of molecules A, B, and we take a linear combination of Euclidean distance and graph geodesic distance. The objective function of GW penalizes large differences between $\mathbf{D}_{ik}^{\mathrm{A}}, \mathbf{D}_{jl}^{\mathrm{B}}$, resulting in transport plans $\mathbf{P}$ that transport a pair of atoms in A to another pair of atoms in B whose distances are similar.  

In this application, there are natural constraints on $\mathbf P$ depending on specific molecular structures and alignment problems. 
For example, atoms have labels such as element names and SYBYL types, and we often require the atom labels to match in the assignment. 
To enforce this constraint, we consider the following fused supervised Gromov-Wasserstein (fsGW) problem \cite{cang2022supervised,cang2024supervised}:
\begin{align}
    \label{elementfsgw}
    \mathop{\arg \min}_{\mathbf P\in \Gamma}\ (1-\alpha)\langle \mathbf C, \mathbf P\rangle_F + \alpha \sum_{i,j,k,l}(\mathbf D^{\mathrm{A}}_{ik} - \mathbf D^{\mathrm{B}}_{jl})^2\mathbf P_{ij}\mathbf P_{kl}
\end{align}where $\alpha \in [0,1]$ is a weight parameter balancing the Wasserstein term and Gromov-Wasserstein term, $\langle, \rangle_F$ is the Frobenius product, and $\mathbf C_{ij} = 0$ if the atom $i$ in A and atom $j$ in B are of the same label, and $\mathbf C_{ij} = \infty$ otherwise. Such a cost matrix $\mathbf C$ prohibits label mismatch since a nonzero $\mathbf P_{ij}$ between different labels would induce a infinite transport cost. When $\alpha = 0$, the fsGW becomes a supervised optimal transport (sOT) problem.

Optimal transport problems require the input of marginal distributions $\mathbf{a}, \mathbf{b}$. In this work, we take $\mathbf{a} = \mathbf{b} = \frac{1}{n}\mathbf{1}_n$. Since a transport plan $\mathbf P$ is a probability matrix, we introduce a heuristic procedure to turn it into a permutation matrix $\mathbf P'$ where $\mathbf P'_{ij} = 1$ when $j = \mathop{\arg \max}_k \mathbf P_{ik}$, and we always check if $\mathbf P'$ is a permutation matrix.

\subsection{Aligning Single Molecules and Clusters of Molecules with OTMol}

When A, B are both a single molecule with $n$ atoms, the inputs to the OTMol algorithm \ref{alg:single} are: $X^{\mathrm{A}}, X^{\mathrm{B}} \in \mathbb R^{n\times 3}$, coordinates of atoms of A and B; $T^{\mathrm{A}}, T^{\mathrm{B}}$, atom labels; $B^{\mathrm{A}}, B^{\mathrm{B}}$, adjacency matrices when A, B are seen as graphs. In this work, we consider three types of labels including element name, atom type, and atom connectivity. Distance matrices $\mathbf D^{\mathrm{A}}, \mathbf D^{\mathrm{B}}$ are calculated from $X^{\mathrm{A}}$ and $B^{\mathrm{A}}$, $X^{\mathrm{B}}$ and $B^{\mathrm{B}}$, respectively. More specifically, if we denote the Euclidean distance matrix by $\mathbf D_{E}$ and the geodesic distance matrix by $\mathbf D_G$, then $\mathbf D = (1-c)\mathbf D_E + c \mathbf D_G$, where $c \in [0,1]$ is a weight parameter. The cost matrix is defined by 
\begin{align*}
    \mathbf C_{ij} = \begin{cases}
                0, & T^{\mathrm{A}}_i = T^{\mathrm{B}}_j,\\
                \infty, & T^{\mathrm{A}}_i \neq T^{\mathrm{B}}_j.
                \end{cases}
\end{align*}
For a given set of $\alpha$, denoted by $L_{\alpha}$, we solve an fsGW problem with $\mathbf{C}, \mathbf{D}^{\mathrm{A}}, \mathbf{D}^{\mathrm{B}}$ for each $\alpha$ and choose the atom assignment that has the lowest RMSD. In practice, we often prefer an atom assignment where most bonds are matched. If two bonded atoms in the reference A are not bonded after assignment, we say that this a mismatched bond. We choose the lowest RMSD assignment from the set of atom assignments that have the lowest percentage of mismatched bonds.

\begin{algorithm}
\caption{Alignment of single molecules. }\label{alg:single}
\begin{algorithmic}
\Function{OTMol}{$X^{\mathrm{A}}, X^{\mathrm{B}}, T^{\mathrm{A}}, T^{\mathrm{B}}, B^{\mathrm{A}}, B^{\mathrm{B}}, L_{\alpha}, c$}
\State Calculate $\mathbf C$ from $T^{\mathrm{A}}, T^{\mathrm{B}}$, and $\mathbf D^{\mathrm{A}}=(1-c)\mathbf D^{\mathrm{A}}_E+c\mathbf{D}_G^{\mathrm{A}}, \mathbf D^{\mathrm{B}}=(1-c)\mathbf D^{\mathrm{B}}_E+c\mathbf{D}_G^{\mathrm{B}}$ from $X^{\mathrm{A}}, X^{\mathrm{B}}, B^{\mathrm{A}}, B^{\mathrm{B}}, c$
\For{$\alpha \in L_{\alpha}$}
\State $\mathbf{P} \gets$ \Call{fsGW}{$\mathbf{C}, \mathbf{D}^{\mathrm{A}}, \mathbf{D}^{\mathrm{B}}, \alpha$}
\State Calculate $\mathbf{P'}$ 
\If {$\mathbf{P'}$ is a permutation}
    \State Run Kabsch algorithm for $X^{\mathrm{A}}, X^{\mathrm{B}}, \mathbf{P}'$ and get $\mathrm{RMSD}_{\alpha}$
\EndIf
\EndFor
\State Take the lowest $\mathrm{RMSD}_{\alpha}$.
\EndFunction
\end{algorithmic}
\end{algorithm}
When A, B are complexes of several different molecules, we can modify the atom labels and use algorithm \ref{alg:single} to ensure that the assignment is between the same type of molecules. 

When A, B are clusters of individual atoms of the same element, such as noble gas clusters, the cost matrix $\mathbf C$ is not needed. For a given set of $p$, denoted by $L_{p}$, the OTMol algorithm \ref{alg:noblegas} will first solve a GW problem with inputs $\mathbf{D}^{\mathrm{A}}=(\mathbf{D}^{\mathrm{A}}_E)^{\circ p}, \mathbf{D}^{\mathrm{B}}=(\mathbf{D}^{\mathrm{B}}_E)^{\circ p}$ (Hadamard power) for each $p$ and obtain the aligned coordinates $\mathbf{T}(X^{\mathrm{B}})$.
The assignment from the GW will be refined by solving a Kantorovich optimal transport (OT) problem 
\begin{align}
    \mathop{\arg \min}_{\mathbf{P} \in \Gamma}\ \langle \mathbf{P}, \mathbf{M} \rangle_{F}
\end{align}where $\mathbf{M}_{ij} = \|\mathbf{x}^{\mathrm{A}}_i-\mathbf{T}(\mathbf{x}^{\mathrm{B}}_j)\|_2^2$. Then we choose the atom assignment that has the lowest RMSD. 

\begin{algorithm}
\caption{Alignment of clusters of individual atoms of the same element.}\label{alg:noblegas}
\begin{algorithmic}
\Function{OTMol}{$X^{\mathrm{A}}, X^{\mathrm{B}}, L_p$}
\For{$p \in L_p$}
\State Calculate $\mathbf{D}^{\mathrm{A}}=(\mathbf{D}^{\mathrm{A}}_E)^{\circ p}, \mathbf{D}^{\mathrm{B}}=(\mathbf{D}^{\mathrm{B}}_E)^{\circ p}$ from $X^{\mathrm{A}}, X^{\mathrm{B}}$
\State $\mathbf{P} \gets$ \Call{GW}{$\mathbf{D}^{\mathrm{A}}, \mathbf{D}^{\mathrm{B}}$}
\State Calculate $\mathbf{P'}$
\If {$\mathbf{P'}$ is a permutation}
    \State Run Kabsch algorithm for $X^{\mathrm{A}}, X^{\mathrm{B}}, \mathbf{P'}$ and get $\mathbf{T}(X^{\mathrm{B}})$
    \State Calculate $\mathbf{M}$ from $X^{\mathrm{A}}, \mathbf{T}(X^{\mathrm{B}})$
    \State $\mathbf{P} \gets$ \Call{OT}{$\mathbf{M}$}
    \State Calculate $\mathbf{P'}$ 
    \If {$\mathbf{P'}$ is a permutation} 
        \State Run Kabsch algorithm for $X^{\mathrm{A}}, X^{\mathrm{B}}, \mathbf{P}'$ and get $\mathrm{RMSD}_p$
    \EndIf
\EndIf
\EndFor
\State Take the lowest $\mathrm{RMSD}_{p}$.
\EndFunction
\end{algorithmic}
\end{algorithm}

When A, B are clusters of the same molecule, a proper assignment between them should map a molecule to another molecule, without splitting one. 
The OTMol algorithm \ref{alg:wc} first finds an assignment at the molecule level and then finds an assignment at the atom level. 
For each molecule, we take a certain coordinate such as the center of mass and obtain an assignment of all representative coordinates using the OTMol algorithm \ref{alg:perturbation}. 
This algorithm searches for representative coordinate assignments by first adding Gaussian noise to the representative coordinates and then solving Kantorovich OT problems, generating a large collection of suboptimal molecule level assignments. This sampling of suboptimal assignments at the molecule level is crucial, as the optimal atom-level assignments may not arise from the optimal molecule-level assignment. Taking a pair of water clusters as an example, suppose that we use centroids as representative coordinates, the orientation of water molecules is ignored when searching for centroid assignments, and an optimal centroid assignment does not necessarily guarantee an optimal assignment of oxygen and hydrogen atoms, especially when the structures are highly symmetric such as 10-PP1 and 10-PP2. The assignment between representative coordinates will be viewed as the assignment between molecules. OTMol then solves an sOT problem with a block cost matrix $\mathbf{M}$, where only the distances between the same element in the matched molecules are finite. This ensures that atoms of a molecule are not assigned to more than one molecule. 

\begin{algorithm}
\caption{Alignment of clusters of a certain molecule.}\label{alg:wc}
\begin{algorithmic}
\Function{OTMol}{$X^{\mathrm{A}}, X^{\mathrm{B}}, T^{\mathrm{A}}, T^{\mathrm{B}}, l$}
\State Calculate representative coordinates $R^A, R^B$
\State $L \gets$ \Call{PerturbationBeforeGW}{$R^{\mathrm{A}}, R^{\mathrm{B}}, l$}
\For{$\mathbf P' \in L$}
\State Run the Kabsch algorithm for $R^{\mathrm{A}}, R^{\mathrm{B}}, \mathbf P'$ and obtain $\mathbf{T}$ 
\State Calculate $\mathbf{M}$ from $X^{\mathrm{A}}, \mathbf{T}(X^{\mathrm{B}}), T^{\mathrm{A}}, T^{\mathrm{B}}$, and the assignment of molecules $\mathbf{P'}$
\State $\mathbf{P} \gets$ \Call{sOT}{$\mathbf{M}$}
\State Calulate $\mathbf{P'}$
\If {$\mathbf{P'}$ is a permutation}
    \State Run Kabsch algorithm for $X^{\mathrm{A}}, X^{\mathrm{B}}, \mathbf{P}'$ and calculate $\mathrm{RMSD}_{\mathbf{P}'}$
\EndIf
\EndFor
\State Take the lowest $\mathrm{RMSD}_{\mathbf P'}$
\EndFunction
\end{algorithmic}
\end{algorithm}

\begin{algorithm}
\caption{The function PerturbationBeforeGW.}\label{alg:perturbation}
\begin{algorithmic}
\Function{PerturbationBeforeGW}{$X^{\mathrm{A}}, X^{\mathrm{B}}, l$}
\State Initialize an empty list $L$
\For{$i \in \{0,1,\dots,l-1\}$}
\State Add Gaussian noise to $X^{\mathrm{A}}, X^{\mathrm{B}}$ and get $\tilde{X}^{\mathrm{A}}, \tilde{X}^{\mathrm{B}}$
\State Calculate $\mathbf D^{\mathrm{A}}=\mathbf D^{\mathrm{A}}_E, \mathbf D^{\mathrm{B}}=\mathbf D^{\mathrm{B}}_E$ from $\tilde{X}^{\mathrm{A}}, \tilde{X}^{\mathrm{B}}$
\State $\mathbf{P}_{\mathrm{GW}} \gets$ \Call{GW}{$\mathbf{D}^{\mathrm{A}}, \mathbf{D}^{\mathrm{B}}$}
\State Calculate $\mathbf{P'}_{\mathrm{GW}}$
\If {$\mathbf{P'}_{\mathrm{GW}}$ is a permutation}
    \State Run Kabsch algorithm for $X^{\mathrm{A}}, X^{\mathrm{B}}, \mathbf{P'}_{\mathrm{GW}}$ and get $\mathbf{T}(X^{\mathrm{B}})$
    \State Calculate $\mathbf{M}$ from $X^{\mathrm{A}}, \mathbf{T}(X^{\mathrm{B}})$
    \State $\mathbf{P} \gets$ \Call{OT}{$\mathbf{M}$} 
    \State Calculate $\mathbf{P'}$
    \If {$\mathbf{P'}$ is a permutation}
    \State Append $\mathbf{P'}$ to $L$
    \EndIf
\EndIf
\EndFor
\State Return $L$
\EndFunction
\end{algorithmic}
\end{algorithm}

In the special case of aligning molecules, the feasible transport plans always fully satisfy the exact marginal distribution constraints in conventional optimal transport settings. Therefore, in practice, the sOT and fsGW problems outlined above can also be solved using existing solvers for the conventional OT \cite{bonneel2011displacement}, GW \cite{peyre2016gromov}, and fGW (fused Gromov-Wasserstein) problems \cite{titouan2019optimal}, by replacing the $\infty$ entries in the cost matrix $\mathbf{C}$ with a large constant.

\subsection{Biomedical Molecular Datasets}

To evaluate molecular alignment performance in biologically and pharmaceutically relevant contexts, we assembled a benchmark that integrates curated biomedical datasets alongside established benchmarks from the ArbAlign study. The curated datasets encompass diverse biomolecular systems, including drug molecules, lipids, nucleic acids, and water clusters, chosen for their biomedical significance and their range of flexibility, stereochemistry, and conformational diversity. The ArbAlign datasets contribute noble gas clusters, small peptides, and heterogeneous atmospheric clusters, serving as robust baselines for permutation-sensitive alignment tasks. Together, these datasets form a rigorous and multifaceted platform to benchmark alignment algorithms for applications in molecular recognition, drug design, and structural bioinformatics.

\textbf{ATP dataset}
This dataset included three types of ATP conformers: (i) a protein-bound ATP conformer extracted from a Protein Data Bank (PDB) structure (PDB ID: 6LN3), (ii) a protein-free ATP conformer coordinated only to metal ions and water molecules from the Cambridge Structural Database (CSD Entry: GAYLOH), and (iii) ATP conformers extracted from protein--ligand complexes generated by Boltz-2. ATP was selected as a representative small, flexible biomolecule that adopts distinct conformations in different binding environments, making it suitable for assessing alignment robustness.

\textbf{Imatinib dataset}
This dataset comprised: (i) a protein-bound Imatinib conformer from the PDB(PDB ID: 4CSV) , (ii) a protein-free Imatinib conformer from the CSD(CSD Entry: AJIGUZ), and (iii) Imatinib poses extracted from Boltz-2--predicted protein--Imatinib complexes. Imatinib, a tyrosine kinase inhibitor with planar aromatic and heterocyclic rings connected by flexible linkers, presents both rigid and flexible structural features, which can challenge accurate atom mapping during alignment.

\textbf{Cyclic peptide dataset}
This dataset contained two subsets of cyclic peptide pairs from the CSD: (i) five pairs with small backbone $\phi$--$\psi$ dihedral angle differences (overall average $<$~15$^\circ$; mean: 6.33$^\circ$, min: 4.03$^\circ$, max: 12.10$^\circ$), representing highly similar macrocyclic scaffolds, and (ii) five pairs with large backbone $\phi$--$\psi$ dihedral angle differences (overall average $>$~60$^\circ$; mean: 85.58$^\circ$, min: 64.32$^\circ$, max: 96.26$^\circ$), representing more structurally diverse conformations. (iii) The third subset includes a different cyclic peptide pair for each length (4-mer to 8-mer) that exhibits a lower average $\phi$--$\psi$ angle difference (mean: 72.40$^\circ$, min: 52.35$^\circ$, max: 93.45$^\circ$) than those in (ii), but a higher RMSD value according to ArbAlign. This dataset was designed to assess alignment accuracy in macrocyclic systems with varying backbone flexibility. We refer to these three subsets as the similar-dihedral set (small $\phi$--$\psi$ differences), the dissimilar-dihedral set (large $\phi$--$\psi$ differences), and the angle--RMSD discordant set (smller dihedral variation than (ii) but higher global RMSD given by ArbAlign).

\textbf{Phospholipid (DLP) dataset}
This dataset focuses on a phospholipid conformer, dilauroylphosphatidylcholine (DLP), extracted from a protein--ligand complex structure (PDB ID: 1LN1). The conformer was generated by Boltz-2 based on the docked pose within the complex. Phospholipids like DLP contain long aliphatic chains and polar headgroups, presenting challenges in alignment due to their extended and amphipathic nature.

\textbf{Unsaturated fatty acid (EIC) dataset}
The EIC dataset includes a single conformer of the unsaturated fatty acid extracted from a protein--ligand complex (PDB ID: 5BVS), generated by Boltz-2. Unsaturated fatty acids like EIC possess flexible hydrocarbon tails with cis double bonds, making them prone to conformational diversity that can affect alignment accuracy.

\textbf{Sugar (BGC and BGC\_GLC) dataset}
This dataset contains two sugar conformers: (i) BGC, a mannose-derived monosaccharide, and (ii) BGC\_GLC, a disaccharide composed of BGC and glucose. Both conformers were extracted from the same protein--ligand complex (PDB ID: 8W4X), with BGC on chain B and BGC\_GLC on chain C. The conformers were generated using Boltz-2. These molecules highlight the challenge of aligning carbohydrate structures, especially due to their stereochemical complexity and flexibility around glycosidic bonds.

\textbf{DNA (215D) dataset}
This dataset includes a single-stranded DNA conformer extracted from the PDB structure 215D. The conformer was generated using AlphaFold3 (AF3) due to limitations in generating correct DNA conformations with Boltz-2 or RDKit. The DNA chain is modeled as a single continuous strand with one predicted 5' phosphate overhang. This case represents nucleic acid alignment, where backbone continuity and phosphate positioning can affect the comparison.

\textbf{Water cluster datasets}
We evaluate alignment performance on three water cluster datasets.
The first dataset is taken directly from the ArbAlign paper and consists of selected water cluster conformer pairs used in their benchmark evaluations.
The second and third datasets are curated in this work using structures from the Database of Water Cluster Minima \cite{rakshit2019atlas}, which provides optimized geometries for water clusters ranging in size from $n = 3$ to $30$.
In the second dataset, for each cluster size $n$, we select a pair of conformers with the lowest and second-lowest energies among the reported minima. This set represents energy-proximal conformations and is used to assess alignment accuracy in subtle structural variations.
In the third dataset, for each cluster size $n$, we identify the conformer pair with the largest RMSD among the up to 20 lowest-energy structures. This set emphasizes geometric diversity while maintaining energetic plausibility, providing a more challenging test for alignment algorithms.
Together, these datasets support a comprehensive evaluation of alignment robustness across both energetically and geometrically diverse water cluster structures.

\begin{table}[h!]
\small
\centering
\caption{Summary of Datasets Used for Molecular Alignment Evaluation}
\begin{tabular}{|p{2cm}|p{2cm}|p{2cm}|p{8cm}|}
\hline
\textbf{Dataset} & \textbf{Molecular Type} & \textbf{Source} & \textbf{Description}\\
\hline
ATP & Nucleotide & Curated in this work & Protein-bound, metal-coordinated, and Boltz-2-generated ATP conformers. Tests robustness across conformational diversity. \\
\hline
Imatinib & Small molecule drug & Curated in this work & Protein-bound, unbound(CSD Entry: AJIGUZ), and Boltz-2-generated conformers of a flexible kinase inhibitor. Challenges atom mapping due to mixed rigidity/flexibility.\\
\hline
Cyclic peptides & Macrocycles & Curated in this work & Three subsets of five pairs from CSD: one with small ($<$15$^\circ$) and one with large ($>$60$^\circ$) $\phi$--$\psi$ differences, and a third with discordant dihedral–RMSD behavior, where global backbone divergence (RMSD) is high despite moderate local angle differences. Evaluates alignment sensitivity to macrocyclic flexibility.\\
\hline
Phospholipid (DLP) & Lipid & Curated in this work & DLP conformer from protein-ligand complex (PDB 1LN1), generated by Boltz-2. Amphipathic with long aliphatic chains. \\
\hline
Unsaturated fatty acid (EIC) & Lipid & Curated in this work & Flexible hydrocarbon chain with cis double bonds from PDB 5BVS. Boltz-2-generated. \\
\hline
Sugar (BGC, BGC\_GLC) & Carbohydrates & Curated in this work & Monosaccharide and disaccharide from PDB 8W4X, Boltz-2-generated. Tests stereochemical and glycosidic bond alignment. \\
\hline
DNA (215D) & Nucleic acid & Curated in this work & Single-stranded DNA conformer generated by AlphaFold3. Assesses backbone continuity and phosphate positioning. The reference structure is collected from PDB database(PDB ID: 215D).\\
\hline
Water cluster (ArbAlign) & Water clusters & ArbAlign paper & Benchmark pairs of water cluster conformers used in the original ArbAlign study. \\
\hline
Water clusters (minima pairs) & Water clusters & Curated in this work & From \cite{rakshit2019atlas}; pairs with lowest and second-lowest energy for $n=3$ to $30$. Tests subtle structural variations. \\
\hline
Water clusters (max RMSD pairs) & Water clusters & Curated in this work & From \cite{rakshit2019atlas}; pairs with largest ArbAlign RMSD among lowest 20 energy structures. Evaluates geometric diversity. \\
\hline
FGG Tripeptide & Peptides & ArbAlign paper & 15 FGG conformers (one global minimum + 14 higher-energy). Contains a stereocenter to evaluate chirality-sensitive alignment. \\
\hline
S1-MA-W1 cluster & Atmospheric hydrate & ArbAlign paper & Cluster with bisulfate, methylammonium, and water. Requires precise atom-type and hybridization-aware mapping. \\
\hline
Neon clusters & Noble gas clusters & ArbAlign paper & Neon clusters of sizes $n=$10 to 1000. Permutationally symmetric, tests optimal atom reordering under dispersion interactions. \\
\hline
\end{tabular}
\label{tab:overall_summary_datasets}
\end{table}

\subsection{Evaluation Metrics}

\subsubsection{Bond Connectivity Inconsistency (BCI) Check}

To evaluate the topological fidelity of structural alignments, we implemented a Bond Connectivity Inconsistency (BCI) metric. This measure quantifies the percentage of covalent bonds in the reference structure that are not preserved in the aligned structure, thereby capturing gross mismatches in atom–atom connectivity.
Specifically, all covalent bonds (including both single and double bonds) in the reference molecule are extracted as unordered atom pairs (tuples), ignoring bond order and chemical environment. A bond is considered preserved if the same atom pair is found in the aligned structure. Conversely, if a reference bond is not recovered after alignment, it contributes to the BCI score. Thus, the BCI reflects alignment-induced topological disruption, independent of bond type, valence, or atomic hybridization.
A BCI of 0\textbf{\%} indicates a perfect alignment in which all reference bonds are retained, while higher values correspond to increasing levels of connectivity inconsistency. This topological validation serves as a structural prerequisite for chemically plausible alignments and is particularly useful for detecting severe misalignments that disrupt molecular backbone or ring structures.
Note that BCI is a structural metric and does not assess the chemical correctness of the aligned conformation. It is used as a requirement-level consistency check prior to further chemical or energetic evaluation. The user can let OTMol algorithm \ref{alg:single} choose the lowest RMSD assignment from the set of lowest BCI assignments.

\subsubsection{Atom Mapping Inconsistency (AMI) Check}

In addition to the bond-level Bond Connectivity Inconsistency (BCI), we introduce an atom-level metric termed Atom Mapping Inconsistency (AMI) to assess the chemical fidelity of the alignment between reference and predicted structures for cases. 
AMI quantifies the proportion of atoms whose mapped counterparts fall into chemically inconsistent environments, based on a Maximum Common Substructure (MCS) framework. Specifically, we perform an MCS search using RDKit to identify the largest shared substructure between the reference and aligned molecules. Atom-to-atom correspondences are then extracted from the MCS match.
To assess whether aligned atoms are chemically equivalent, we compute canonical atom ranks for both molecules. These ranks group atoms into equivalence classes based on local chemical environment and molecular symmetry, while ignoring stereochemistry and isotopes. For each matched atom pair, we verify whether the aligned atom belongs to the same symmetry group as the reference atom. Any atom whose mapped counterpart falls outside its expected group is counted as inconsistent.
The AMI score is reported as the percentage of mismatched atoms among all atoms in the MCS. An AMI of 0\% indicates a chemically coherent atom mapping, whereas higher values reflect increasing degrees of chemically implausible alignment at the atomic level.
Unlike BCI, which captures gross topological mismatches, AMI detects more subtle alignment errors involving local environment discrepancies, making it a complementary and chemically sensitive evaluation criterion for some cases requiring more details.

\section{Results}\label{results}

To systematically assess the performance of OTMol in biologically and chemically relevant scenarios, we performed alignment experiments across a suite of curated biomedical molecular datasets. These benchmarks span diverse molecular classes, including peptides, nucleic acids, drug-like compounds, and biomolecular clusters, as well as previously established ArbAlign datasets, ensuring coverage of both biologically significant systems and permutation-sensitive test cases(see Method section). This comprehensive evaluation framework enables us to probe the robustness, chemical fidelity, and computational efficiency of OTMol in contexts relevant to structural biology, medicinal chemistry, and molecular informatics.
 For most datasets, we only used element names as atom labels, with the exception of the FGG and atmospheric hydrates datasets. The OTMol alignments were computed by solving the fsGW and sOT problems using the fGW and Earth Mover's Distance (EMD) solvers from the Python Optimal Transport (POT) library \cite{flamary2021pot, flamary2024pot}. 

\subsection{Performance on Biomedical Molecular Dataset}

\subsubsection{Overall Description}

We used the OTMol algorithm \ref{alg:single} to align eight biomolecules with their conformers. 
In the experiment, we let $L_{\alpha} = \{0, 0.01, 0.02, \dots, 0.99, 1\}$ and $c=1$, meaning that only geodesic distance is used. We only choose the assignment with the lowest RMSD from the set of assignments that have the lowest BCI values. In other words, we hope that the atom assignment is very close to a graph isomorphism.
OTMol is able to match all edges for all pairs of conformers. Although ArbAlign gives lower RMSD values, its atom assignments usually involve inconsistent bond connectivity and atom-mismatching issues. The bond connection condition in the aligned molecules is evaluated with  Bond Connectivity Inconsistency (BCI) and the atom mapping condition within chemical environment is evaluated with Atom Mapping Inconsistency (AMI), see method Evaluation Metrics subsection for more details.

\subsubsection{Application to Adenosine Triphosphate}

\begin{figure}[htbp]
    \centering
    \includegraphics[width=0.7\linewidth]{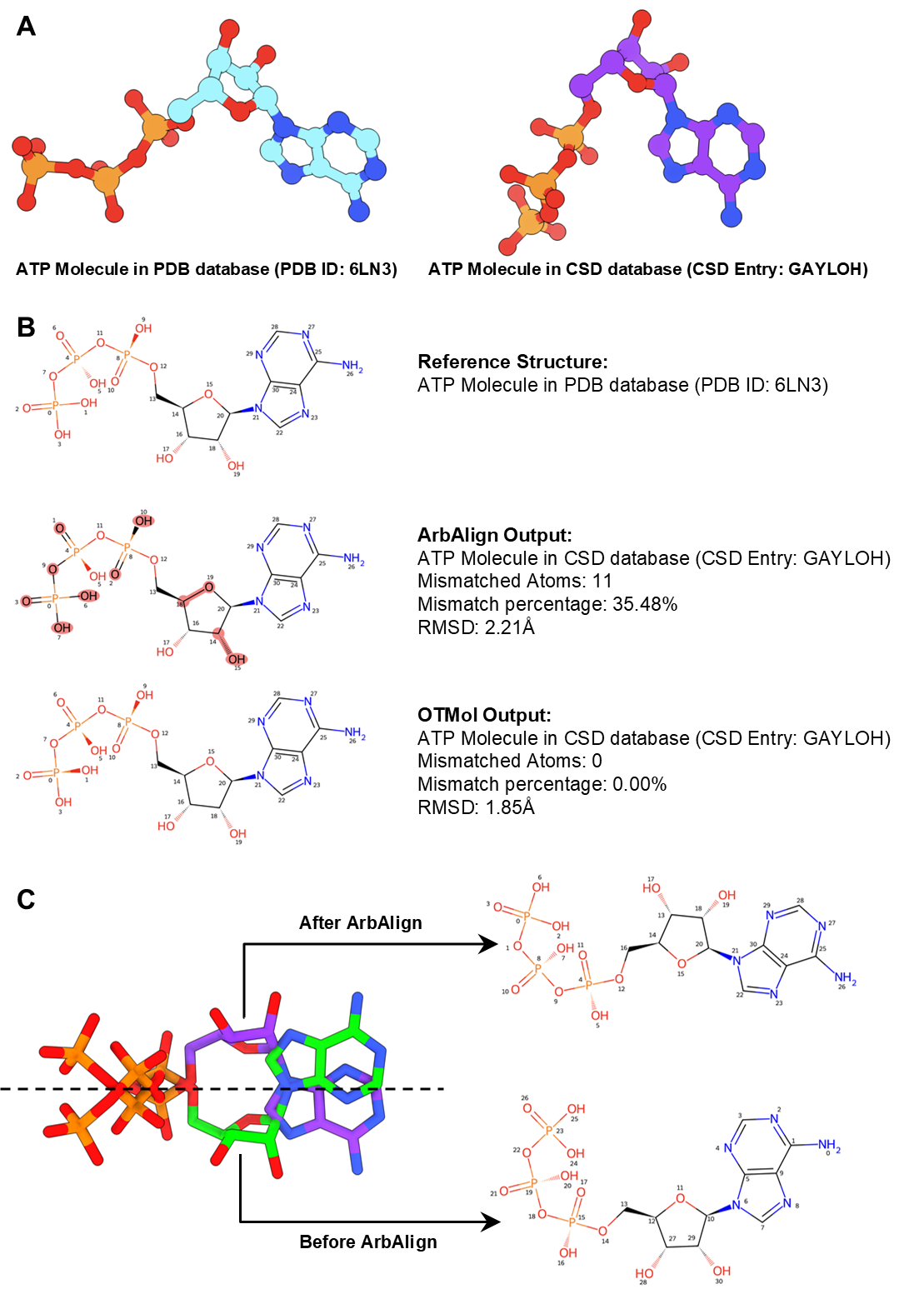}
    \caption{(A) Visualization of the ATP molecules from PDB and CSD databases. Only the ATP part is shown. (Left: ATP molecule in PDB database, PDB ID: 6LN3; Right: ATP molecule in CSD database, CSD Entry: GAYLOH). (B) The atom mismatching analysis of the ATP molecule (collected from the CSD database) aligned by ArbAlign and OTMol, respectively. The mismatched heavy atoms are colored in red. (C) The diagram of the ATP molecule (generated by Boltz-2; model-0) before and after alignment via ArbAlign in both 3-dimensional and 2-dimensional diagrams. (Lower: original ATP molecule configuration before alignment via ArbAlign; Upper: ATP molecule configuration after alignment via ArbAlign.)}
    \label{fig:ATP}
\end{figure}

Adenosine triphosphate (ATP) is essential for numerous cellular processes, serving both as an energy source and a phosphate donor. Due to its structural flexibility, ATP can adopt diverse conformations under different conditions, making structural comparisons both meaningful and challenging \cite{ATP_Hara2015, ATP_Eri_Kobayashi2013}.
Two ATP conformers from different structural databases were selected for evaluation. The conformer from the Protein Data Bank (PDB ID: 6LN3; \autoref{fig:ATP}A, Left) is bound within a receptor pocket, whereas the conformer from the Cambridge Structural Database (CSD Entry: GAYLOH; \autoref{fig:ATP}A, Right) is coordinated only to metal ions and water molecules. In addition, protein--ligand complexes generated by Boltz-2 were used to extract ATP conformers for comparison. Structural files from different sources often differ in atom ordering, naming conventions, geometries, and other formatting aspects, providing effective test cases for assessing alignment capabilities and evaluating result plausibility.

Applying ArbAlign to align ATP conformers from different sources revealed two main issues: (1) frequent atom mismatches in the aligned structures used for RMSD calculation, and (2) changes in stereochemistry when a reflection operation was applied. Accurate atom mapping is essential for meaningful RMSD calculation and interpretation. In one ArbAlign alignment, 9 oxygen atoms and 2 carbon atoms were mismatched, corresponding to an AMI mismatch rate of 35.48\% which is unexpectedly high for ATP (\autoref{fig:ATP}B), a well-characterized and relatively simple molecule with 31 heavy atoms. Moreover, in some cases, the entire molecular chirality was inverted due to reflection. Since ArbAlign applies reflection operations by default, this can yield mirrored molecules that minimize RMSD but distort stereochemistry. For example, aligning a Boltz-generated ATP pose to the reference structure resulted in a mirrored conformation with three reversed chiral centers (\autoref{fig:ATP}C).

\subsubsection{Application to Tyrosine Kinase Inhibitor: Imatinib}
Imatinib mesylate (marketed as Gleevec\textsuperscript{\textregistered} or Glivec\textsuperscript{\textregistered} by Novartis, Switzerland) represents a landmark in rational drug design, as it specifically targets a single molecular event that is both necessary and sufficient to initiate carcinogenesis \cite{Imatinib_DEININGER20052640,Imatinib_1Naveed_Iqbal}.

\begin{figure}[htbp]
    \centering
    \includegraphics[width=0.75\linewidth]{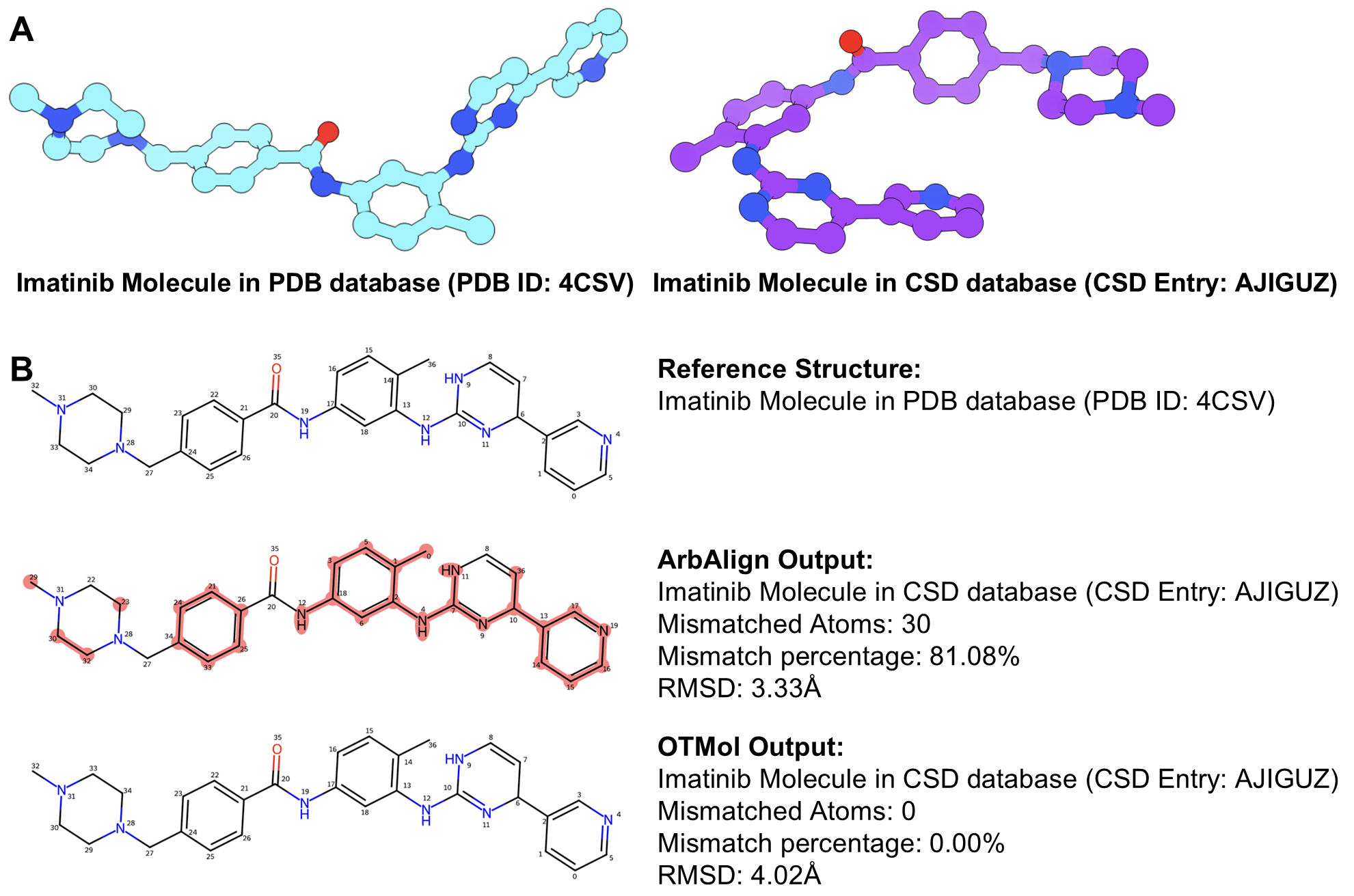}
    \caption{(A) Visualization of the Imatinib molecules from PDB and CSD databases. Only the ATP part is shown. (Left: ATP molecule in PDB database, PDB ID: 4CSV; Right: ATP molecule in CSD database, CSD Entry: AJIGUZ). (B) The atom mismatching analysis of Imatinib molecule (collected from CSD database) aligned by ArbAlign and OTMol, respectively. The mismatched heavy atoms are colored in red.}
    \label{fig:imatinib}
\end{figure}

We performed structural comparison analyses of different Imatinib conformers, including free forms and those bound to either wild-type or mutant platelet-derived growth factor receptors (PDGF-R). Such conformational differences can provide insights into potential binding modes and possible mechanisms of drug resistance in mutant receptors. Similar to the ATP case (\autoref{fig:ATP}), two structures were selected from the Protein Data Bank (protein-bound pose) and the Cambridge Structural Database (protein-free pose) (\autoref{fig:imatinib}A). In addition, protein--Imatinib complexes were predicted by Boltz-2, and the extracted Imatinib poses were analyzed.

Imatinib contains planar aromatic and heterocyclic rings, along with flexible linkers and multiple rotatable bonds, making it a particularly challenging case for alignment. The AMI mismatch percentage reached 81.08\%, with most ring atoms failing to map to the correct or equivalent positions. Although the ArbAlign-generated alignment yielded an RMSD of 3.33~\AA{}, which is lower than that from OTMol, this suggests that ArbAlign may minimize the overall RMSD at the expense of correct atom mapping (Figure \ref{fig:imatinib}B).

\subsubsection{Application to Cyclic Peptide Backbones}

To evaluate structural similarity and conformational accuracy in cyclic peptides, backbone RMSD is a key metric for assessing how well predicted or designed peptide backbones match known, expected, or reference structures. This is particularly important for benchmarking alignment algorithms in applications such as cyclic peptide design and discovery. In this context, the minimal RMSD value serves as an appropriate indicator of overall cyclic peptide scaffold similarity and can provide useful insights for cyclic peptide design.
In our evaluation dataset, we selected three groups of cyclic peptide pairs from the CSD database: five pairs with small backbone $\phi$ and $\psi$ dihedral angle differences (overall average $\phi$--$\psi$ difference $<$~15$^\circ$) and five pairs with large backbone $\phi$ and $\psi$ dihedral angle differences (overall average $\phi$--$\psi$ difference $>$~60$^\circ$), and five angle--RMSD discordant pairs, where the overall average $\phi$--$\psi$ difference is lower than that of the second set but the RMSD values given by ArbAlign are larger. Further details are provided in the Biomedical Molecular Datasets subsection of Materials and Methods Section.

\begin{figure}[htbp]
    \centering
    \includegraphics[width=0.7\linewidth]{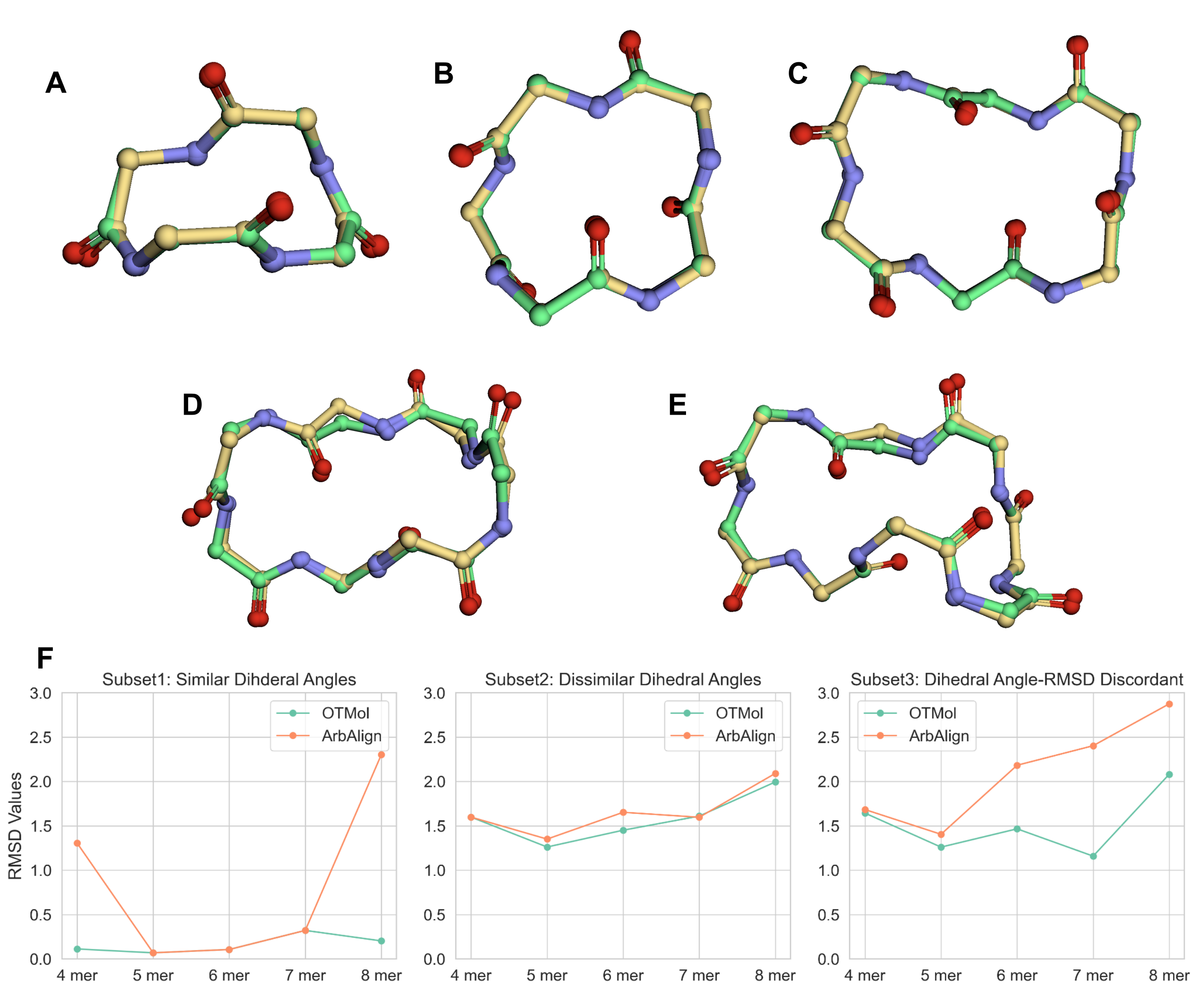}
    \caption{(A, B, C, D, E) The OTMol alignment of the 4-mer pair, 5-mer pair, 6-mer pair, 7-mer pair, and 8-mer pair in the dataset of similar dihedral angle dataset, respectively.
    (F) The RMSD comparison for the three cyclic peptides datasets.}
    \label{otmol_cp}
\end{figure}

The OTMol algorithm \ref{alg:single} was applied to three small datasets of cyclic peptide backbones. For each pair, $L_{\alpha} = \{0, 0.01, 0.02, \dots, 0.99, 1\}$, and $c =0.5$. We only choose the assignment with the lowest RMSD from the set of assignments that have the lowest BCI values.
In theory, a pair of cyclic peptides with similar dihedral angles should have similar structures, so the true RMSD should be very low. OTMol yields very low RMSD (less than 0.5\AA) for all pairs with the dataset of similar dihedral angles (Figure \ref{otmol_cp}). 
The bonds are all recalled in OTMol-aligned structures, i.e., mismatched bond percentage is 0\%, and these results show lower RMSD, indicating the more accurate atom mapping and alignment results. 

\subsubsection{Other Results in Biomedical Molecules}

We also collect and compare other important biological molecules, such as DNA, lipids and sugars, as decribed in the Methods section. These results are consistent with the previous results in ATP and Imatinib. The Bond Connection Inconsistency rate is 0\% in our OTMol aligned results while the mean Bond Connection Inconsistency rate of 24.5~$\pm$~15.03\%  in those ArbAlign results (See  \autoref{tab:mismatched_bond_stats}). The RMSD value is only meaningful under the condition that the mismatched bond percentage is close to 0.

\begin{table}[htbp]
\centering
\caption{Summary statistics of \%Bond Connectivity Inconsistency for generated molecules aligned by ArbAlign in the Biomedical Molecular Dataset. All values are shown as percentages.}
\label{tab:mismatched_bond_stats}
\begin{tabular}{llrrrrrrr}
\toprule
\textbf{PDB ID} & \textbf{Molecule} & \textbf{Mean} & \textbf{Std} & \textbf{Min} & \textbf{Max} & \textbf{Q1} & \textbf{Median} & \textbf{Q3} \\
\midrule
6ln3   & ATP      & 28.46\% & 3.84\%  & 15.15\% & 39.39\% & 27.27\% & 30.30\% & 30.30\% \\
5bvs   & EIC      & 24.74\% & 15.67\% &  0.00\% & 52.63\% & 15.79\% & 18.42\% & 36.84\% \\
8w4x   & BGC      & 16.80\% & 20.55\% &  0.00\% & 70.00\% &  0.00\% &  0.00\% & 30.00\% \\
4csv   & Imatinib& 21.23\% & 16.20\% &  0.00\% & 63.41\% &  8.54\% & 21.95\% & 30.49\% \\
215d   & DNA      &  0.00\% & 0.00\%  &  0.00\% &  0.00\% &  0.00\% &  0.00\% &  0.00\% \\
8w4x   & BGCGLC   & 30.20\% & 17.41\% &  0.00\% & 65.00\% & 20.00\% & 30.00\% & 40.00\% \\
1ln1   & DLP      & 49.74\% & 11.64\% & 28.30\% & 79.25\% & 41.51\% & 50.94\% & 59.91\% \\
\bottomrule
\end{tabular}
\end{table}

\subsection{Performance on ArbAlign Dataset}

In this section, we evaluate OTMol on the benchmarking datasets from ArbAlign.

\subsubsection{Application to Small Peptides FGG and Atmospheric Hydrates}

The FGG dataset consists of tripeptides composed of one phenylalanine (F) and two glycines (G), where phenylalanine contributes an aromatic side chain and a chiral center. We calculated the RMSD between the global minimum structure (FGG\_55) and 14 higher-energy conformations. The presence of a stereocenter in phenylalanine allows us to assess chirality preservation in the alignment process. 
Similarly, the S1-MA-W1 dataset consists of a pair of configurations of an atmospheric cluster containing one bisulfate ion, one methylammonium ion, and one water molecule. For the S1-MA-W1 dataset, we performed alignment between the reference structure s1-ma-w1-1 and other molecules. In contrast to symmetric water or noble gas clusters, where all atoms of the same element have uniform hybridization and connectivity, the atoms in S1-MA-W1 differ significantly in bonding environments. For instance, oxygen atoms exist in both sp\textsuperscript{2} and sp\textsuperscript{3} hybridizations. Assignments based only on atom names can lead to incorrect mappings, such as aligning chemically distinct oxygen atoms. Therefore, accurate alignment in this dataset requires incorporating finer atomic labels, such as SYBYL Mol2 types and atom connectivity.

\begin{figure}
    \centering
    \includegraphics[width=0.7\linewidth]{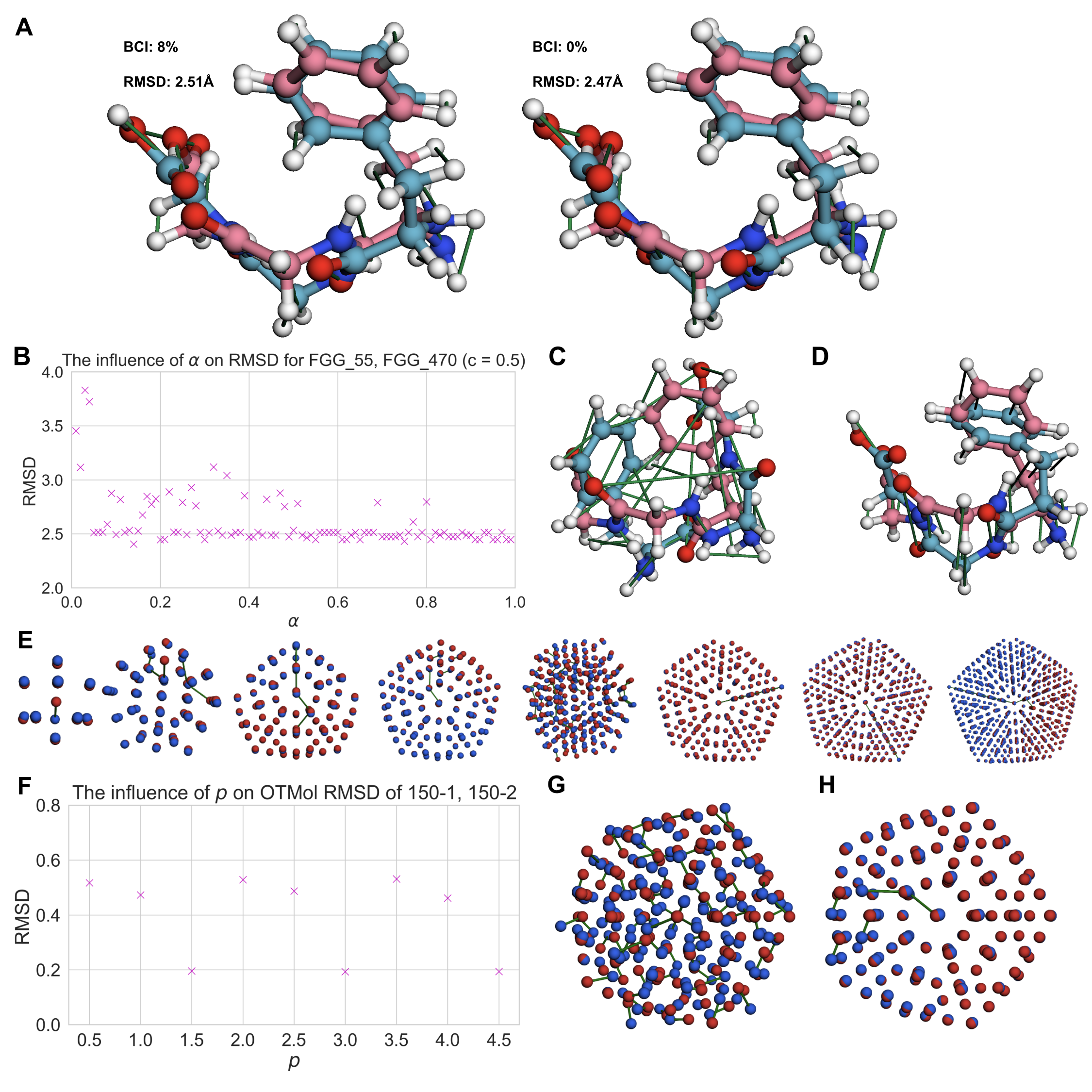} 
    \caption{(A) The OTMol alignment of FGG$\_$55 (reference structure; colored in red) and FGG$\_$470 (structure to be aligned; colored in blue) when $c = 0$ (left) or $0.5$ (right), respectively. Element names are used as labels. (B) The influence of $\alpha$ on OTMol RMSD of the pair FGG$\_$55 (colored in red), FGG$\_$470 (colored in blue) when $c = 0.5$. (C) The OTMol alignment of FGG$\_$55 (colored in red) and FGG$\_$470 (colored in blue) when $\alpha = 0.01, c = 0.5$. RMSD = 3.45\AA. BCI = 92\%. (D) The OTMol alignment of FGG$\_$55 and FGG$\_$470 when $\alpha = 0.62, c = 0.5$. RMSD = 2.45\AA. BCI = 3\%.
    (E) The OTMol alignment of all pairs in the Neon cluster dataset. (F) The influence of $p$ on OTMol RMSD of 150-1, 150-2. (G) The OTMol alignment of 150-1, 150-2 when $p=2$. RMSD = 0.53\AA. (H) The OTMol alignment of 150-1, 150-2 when $p=3$. RMSD = 0.19\AA.}
    \label{fig:fig4}
\end{figure}

Both the tripeptide FGG dataset and the atmospheric hydrates dataset have been previously evaluated using ArbAlign, making them well-suited for comparative benchmarking of alignment performance.
In the experiment, we let $L_{\alpha} = \{0, 0.01, 0.02, \dots, 0.99, 1\}$ for two datasets. We let $c = 0.5$ for the FGG dataset and $c=1$ for the S1-MA-W1 dataset. We only choose the assignment with the lowest RMSD from the set of assignments that have the lowest BCI values. 
The influence of individual $\alpha$ when $c=0.5$ on OTMol RMSD for the pair FGG$\_$55, FGG$\_$470 is shown in Figure \ref{fig:fig4}B.

\subsubsection{Application to Neon Clusters}

Noble gas clusters are held together by weak dispersion forces, enabling them to adopt a vast number of energetically accessible configurations. Rigorous comparison of such clusters requires careful reordering of atoms, as permutations can yield more accurate alignments and lower RMSD values. We employed our OTMol algorithm \ref{alg:noblegas} on the Neon cluster dataset from the ArbAlign paper. The dataset consists of Neon clusters with sizes $n$ = 10, 50, 100, 150, 200, 300, 500, and 1000. In the experiment, we let $L_p = \{0.5, 1,1.5,2,2.5,3,3.5,4,4.5\}$. The OTMol alignment of each pair is shown in Figure \ref{fig:fig4}E. OTMol can align all pairs very closely except for the pair with 200 atoms. The influence of $p$ on the OTMol alignment of 150-1 and 150-2 is shown in \autoref{fig:fig4}F-H. OTMol RMSD value doesn't decrease when $p$ increases.

\begin{figure}[htbp]
    \centering
    \includegraphics[width=0.95\linewidth]{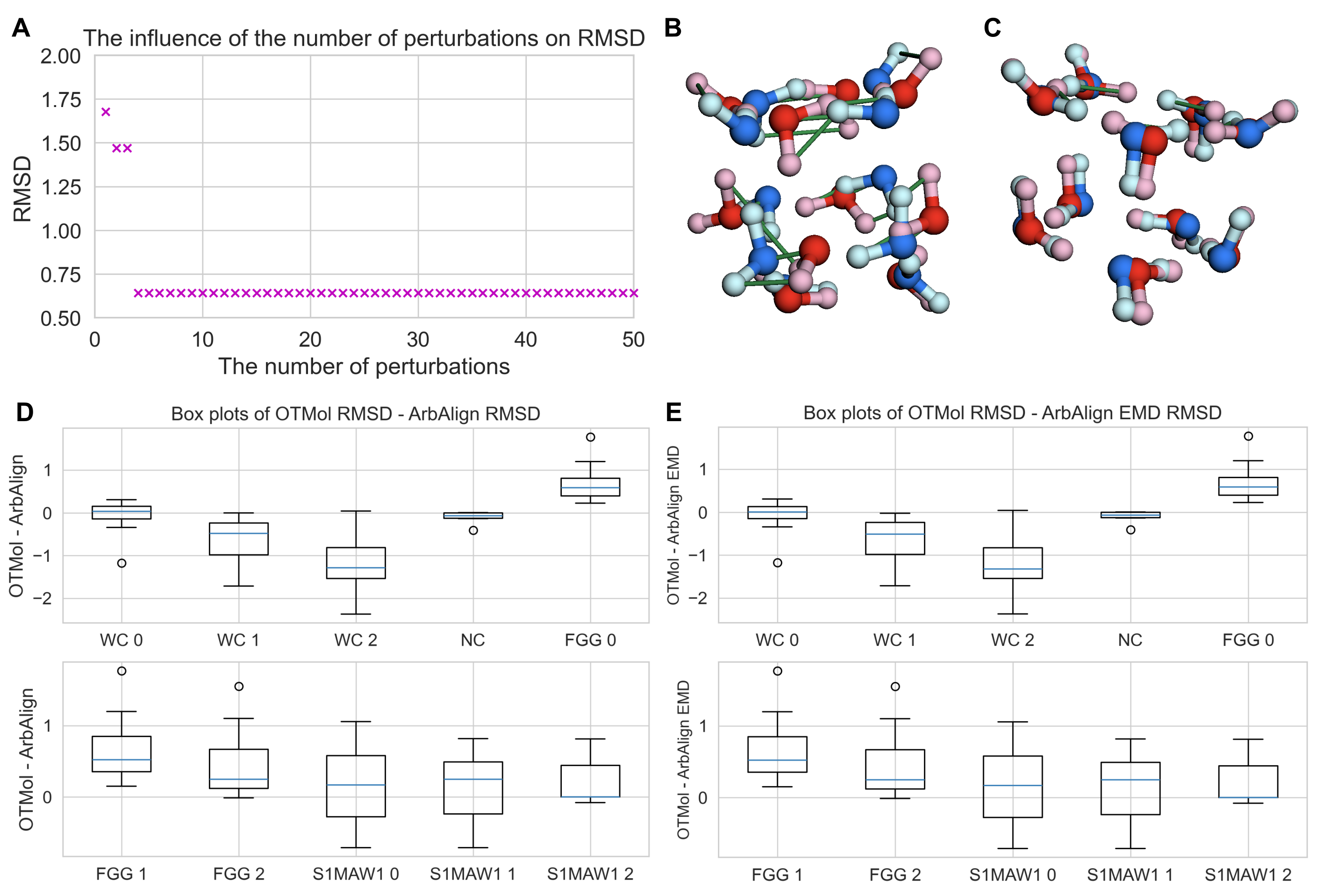}
    \caption{(A) The influence of $l$ on OTMol RMSD of the pair 10-PP1 and 10-PP2. The OTMol RMSD decreases as the number of perturbations increases. (B) The OTMol alignment of 10-PP1(pink: hydrogen atoms; red: oxygen atoms), 10-PP2(light blue: hydrogen atoms; dark blue: oxygen atoms) when $l=2$. RMSD = 1.47\AA. (C) The OTMol alignment of 10-PP1(pink: hydrogen atoms; red: oxygen atoms) , 10-PP2(light blue: hydrogen atoms; dark blue: oxygen atoms)  when $l=21$. RMSD = 0.64\AA. (D) The boxplots of the difference OTMol RMSD - ArbAlign RMSD for all datasets. WC0: the water cluster dataset from the ArbAlign paper; WC1: pairs of water clusters that have the lowest and second-lowest energies for each group of size $n$; WC2: pairs of water clusters that have the largest ArbAlign RMSD for each group of size $n$; NC: Neon clusters;
    FGG: tripeptides FGG (0, 1, and 2 represent three settings of atom labels: element name, SYBYL type, and atom connectivity, respectively); S1MAW1: atmospheric hydrates (0, 1, and 2 represent three settings of atom labels: element name, SYBYL type, and atom connectivity, respectively). (E) The boxplots of the difference OTMol RMSD - ArbAlign EMD RMSD for all datasets. The results of ArbAlign EMD are almost identical to those of ArbAlign.}\label{fig:fig5}
\end{figure}

\subsubsection{Application to Water Clusters}

Water clusters play a critical role in both condensed matter systems \cite{water_cluster_condenseacs.langmuir.1c01322} and biological environments \cite{water_clujster_bio_Pouliquen}. Due to the interchangeability of water molecules within these clusters, it is often necessary to permute monomers, or their constituent oxygen and hydrogen atoms, to achieve optimal alignment prior to RMSD calculation. We compared the results of the OTMol algorithm \ref{alg:wc} and ArbAlign on three water cluster datasets. The first dataset is taken from the ArbAlign paper. The second and third datasets contain water clusters selected from the Database of Water Cluster Minima \cite{rakshit2019atlas} that contains water clusters of sizes $n=3-30$. The second dataset consists of pairs of water clusters that have the lowest and second-lowest energies for each group of size $n$. The third dataset consists of pairs of water clusters that have the largest ArbAlign RMSD for each group of size $n$. We choose the centroid of each molecule as its representative coordinate. The number of perturbations $l$ on the centroid coordinates is 100 for all pairs. The influence of $l$ on OTMol RMSD of the pair 10-PP1 and 10-PP2 is shown in Figure \ref{fig:fig5}A-C.

\FloatBarrier

\subsubsection{Comparison with ArbAlign Results}

\begin{figure}[htbp]
    \centering
    \includegraphics[width=\linewidth]{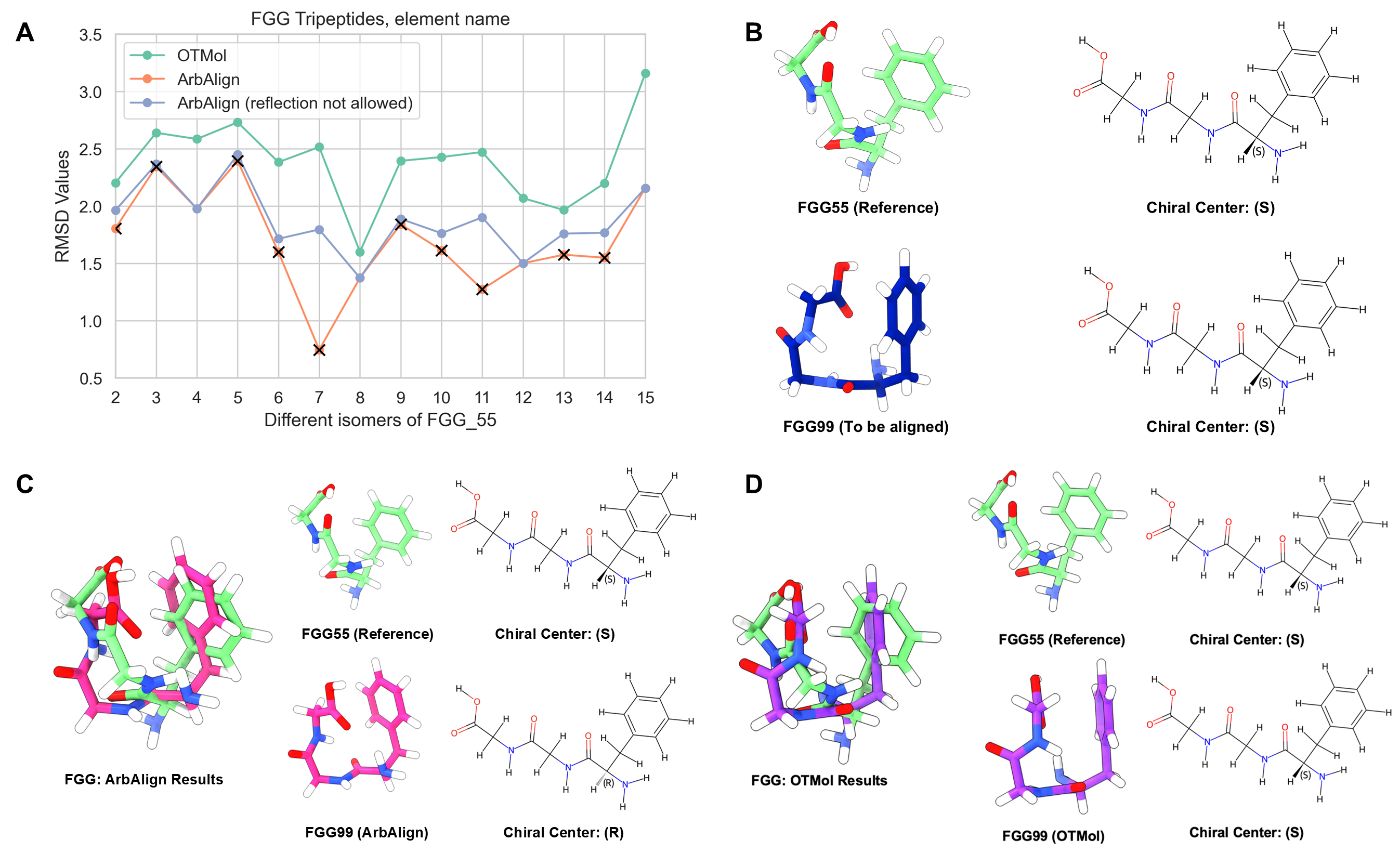}
    \caption{Comparison of OTMol and ArbAlign alignments on the FGG tripeptides. (A) RMSD comparison. A black cross indicates an ArbAlign alignment that involves reflection. We also modified ArbAlign so that reflection is not allowed and calculated results. (B) Visualization of the short peptides FGG55 and FGG99 conformation and their chiral configuration at the phenylalanine(F). (C) Visualization of the short peptides FGG55 and FGG99 conformation aligned via ArbAlign and their chiral configuration at the phenylalanine(F). (D) Visualization of the short peptides FGG55 and FGG99 conformation aligned via OTMol and their chiral configuration at the phenylalanine(F).}
    \label{fig:fgg-alignment}
\end{figure}

\begin{figure}[htbp]
    \centering
    \includegraphics[width=\linewidth]{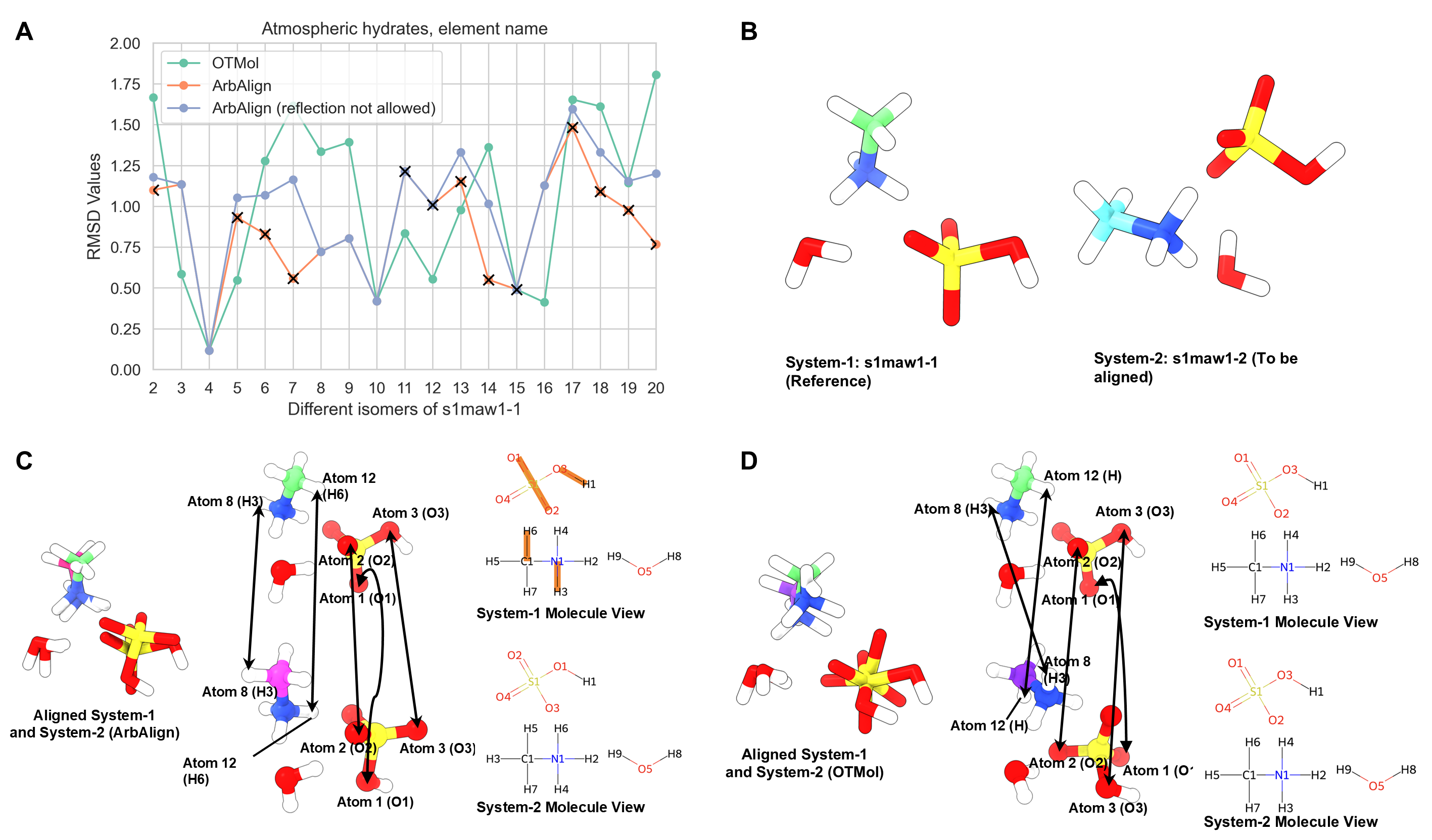}
    \caption{Comparison of alignment results for atmospheric hydrate configurations. (A) RMSD comparison. A black cross indicates an ArbAlign alignment that involves reflection. We also modified ArbAlign so that reflection is not allowed and calculated results. (B) A Visualization of the system configuration (bisulfate ion-methylammonium ion-water molecule) of system-1 (s1maw1-1) and system-2 (s1maw1-2). (C) Anaylisis of the system configuration (bisulfate ion-methylammonium ion-water molecule) of system-1 (s1maw1-1) and system-2 (s1maw1-2) after alignment via ArbAlign. The atom mapping in the reference system (system-1) and the aligned system (system-2) are shown with atom labels and black arrows. The corresponding bonds missing in the system-2 are highlighted in the system-1 in orange color. (D) Anaylisis of the system configuration (bisulfate ion-methylammonium ion-water molecule) of system-1 (s1maw1-1) and system-2 (s1maw1-2) after alignment via OTMol. The atom mapping in the reference system (system-1) and the aligned system (system-2) are shown with atom labels and black arrows. The corresponding bonds missing in the system-2 are highlighted in the system-1 in orange color. No highlighted bond in orange color means there are no mismatched bonds.}
    \label{fig:s1maw-alignment}
\end{figure}

We presents a boxplot analysis of the difference between OTMol RMSD and ArbAlign RMSD across all datasets (Figure \ref{fig:fig5}D). The results demonstrate that OTMol achieves lower RMSD values than ArbAlign for the majority of molecule pairs. We also compared OTMol with another method ArbAlign EMD (Figure \ref{fig:fig5}E). ArbAlign EMD is identical to ArbAlign except that the Hungarian algorithm solver is substituted by the EMD solver in the Python Optimal Transport (POT) library. The RMSD values produced by OTMol on the FGG and S1-MA-W1 datasets are relatively higher than those from ArbAlign. This discrepancy is likely due to the fact that ArbAlign allows reflection, which can alter molecular chirality, whereas OTMol preserves stereochemistry by design. Therefore, RMSD alone should not be the sole criterion for evaluating alignment quality; it is equally important to consider whether the chemical integrity, such as chirality and bonding, is maintained in the aligned structures (see Figure~\ref{fig:fgg-alignment}, \ref{fig:s1maw-alignment}).

\begin{figure}[htbp]
    \centering
    \includegraphics[width=0.95\linewidth]{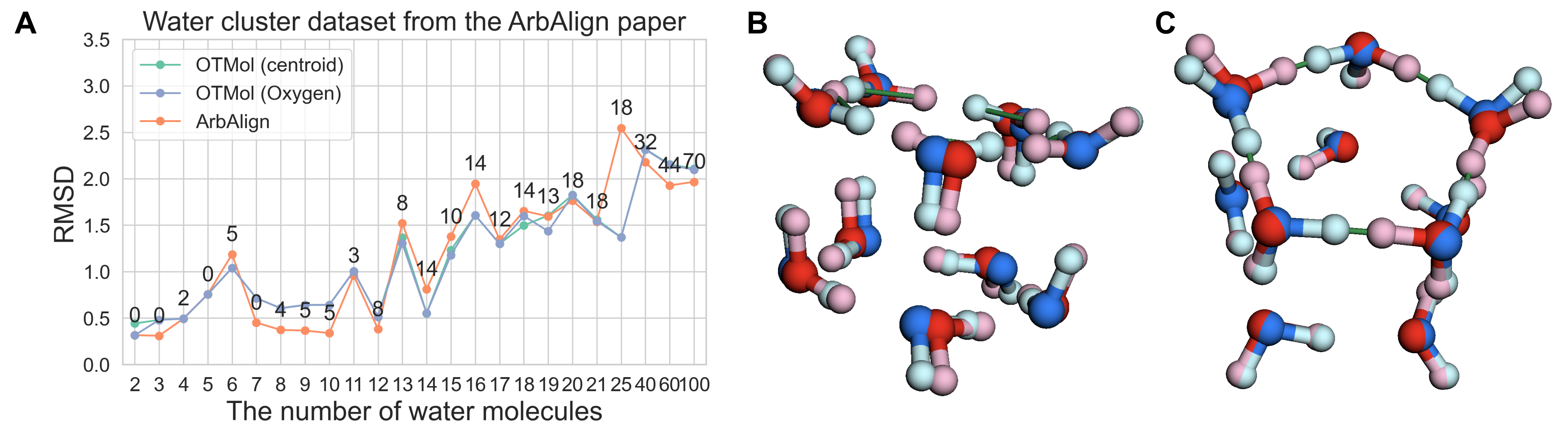}
    \caption{(A) The RMSD values of ArbAlign and OTMol for water cluster datasets. Two representative options, centroid and oxygen atoms, are used. The $x$ axis represents the number of water molecules in clusters. The number above the red dots are the number of mismatched water molecules in ArbAlign alignments. (B, C) Comparison of water cluster alignments between 10-PP1 (pink: hydrogen atoms; red: oxygen atoms)  and 10-PP2 (light blue: hydrogen atoms; dark blue: oxygen atoms)  using OTMol and ArbAlign. (B) is the OTMol alignment and (C) is the ArbAlign alignment. Notably, the ArbAlign alignment exhibits five mismatched water molecules.}
    \label{fig:wc}         
\end{figure}

One limitation of ArbAlign is that it does not guarantee that atoms of a water molecule are assigned to only one water molecule. The number of mismatched water molecules in ArbAlign alignment increases as the number of water molecules in the clusters increases (Figure \ref{fig:wc}). When ArbAlign yields lower RMSD values than OTMol, there are usually mismatched water molecules.

\subsection{Time Advantage}

\begin{figure}
    \centering
    \includegraphics[width=0.95\linewidth]{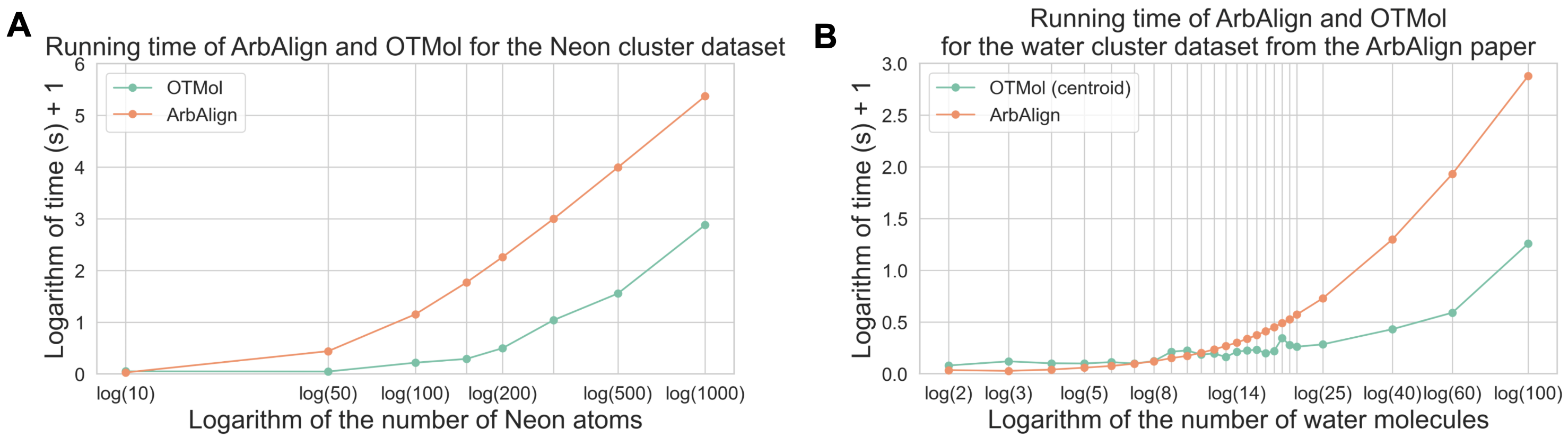}
    \caption{(A, B) The running time comparison of OTMol and ArbAlign for the Neon cluster dataset and the water cluster dataset from the ArbAlign paper.}
    \label{fig:time}
\end{figure}

Running time comparisons between OTMol and ArbAlign shown in Figure \ref{fig:time} were performed on a MacBook Pro equipped with an Apple M1 Pro chip to ensure consistent benchmarking conditions. All three OTMol algorithms \ref{alg:single}, \ref{alg:noblegas}, and \ref{alg:wc} have a theoretical time scaling of $\mathrm{O}(n^3)$. OTMol is much faster than ArbAlign for large systems.
Both OTMol and ArbAlign show increasing computational cost with system size; however, OTMol demonstrates significantly better scalability. While ArbAlign's runtime grows steeply beyond 100 atoms, OTMol maintains noticeably lower execution times, particularly for clusters with 200 or more atoms.

\FloatBarrier

\section{Discussion and Conclusions}\label{discussion}

\subsection{Accurate Atom Mapping and Chirality with OTMol}

Usually, for RMSD calculation algorithms performance comparison, a lower RMSD will be considered as a more accurate RMSD calculation, because the wrong alignment (translate, rotate, swap, etc.) will probably lead to a larger RMSD. However, if the atom mapping algorithm ignores the bonds or chemical environment, such as ArbAlign, the final alignment with wrong mapping will lead to a much lower RMSD than theoretically minimal RMSD value, as mentioned in the Results section, in the case of ATP and Imatinib (\autoref{fig:ATP}, \ref{fig:imatinib}). 
In addition, although ArbAlign does not allow reflection when using the Kabsch algorithm, some initial transformations involve reflections and will change the chirality as mentioned in ATP and peptide FGG alignment cases in the results section (\autoref{fig:ATP}, \ref{fig:fgg-alignment}). The molecule with different  (R) and (S) chiral properties can be totally regarded as a different molecule. Also, for water clusters, sometimes the oxygen and two hydrogen atoms of a certain water molecules are mapped to other water molecules respectively (\autoref{fig:wc}).

Only under the condition that the molecule's chemical information is ensured, a lower RMSD value will suggest a better alignment for a certain RMSD algorithm. Thus, when evaluating these cases, we explore the atom matching situation or bond connection status, in addition to the RMSD value comparison. This comprehensive comparison will be helpful to a better understanding of the alignment-based RMSD calculation. 
The incorporation of bonding information and graph geodesic distance is the main reason why OTMol gives a more correct atom assignment. OTMol also keeps chirality since reflection is not allowed in the use of the Kabsch algorithm.

\subsection{Computational Efficiency with OTMol}

{\autoref{fig:time} compares the runtime of OTMol algorithms 2 and 3, which scale as $\mathrm{O}(n^3)$.} OTMol has an obvious advantage over ArbAlign in computational efficiency for larger molecules or clusters. The running time of OTMol can be controlled by tuning various hyperparameters. In contrast, ArbAlign uses 48 initial spatial transformations, and some transformations are redundant. With the increasing interest and research of macromolecules such as peptides, DNAs, and polysaccharides, and the awareness of the importance of water clusters in biological systems,  a more efficient RMSD calculation algorithm will be preferred.

\subsection{The Fused Supervised Gromov-Wasserstein problem}

The development of OTMol represents a significant advancement in molecular alignment through the application of optimal transport theory. OTMol formulates the molecular alignment problem as a special case (\ref{elementfsgw}) of the general fused supervised Gromov-Wasserstein (fsGW) problem, which is a bilevel optimization
\begin{align}
    \label{}
    \mathop{\max}_{s\in [0, \min\{\|\mathbf{a}\|_1, \|\mathbf{b}\|_1\}]}\mathop{\min}_{\mathbf P\in \mathbf{U}^s(\leq \mathbf{a}, \leq \mathbf{b})} (1-\alpha)\langle \mathbf C, \mathbf P\rangle_F + \alpha \sum_{i,j,k,l}\mathcal{M}(\mathbf D^{\mathrm{A}}, \mathbf D^{\mathrm{B}})_{ijkl}\mathbf P_{ij}\mathbf P_{kl}
\end{align}
where $\mathbf{a}, \mathbf{b}$ are two histograms, $\mathbf{U}^s(\leq\mathbf{a}, \leq\mathbf{b}) = \{\mathbf{P}\in \mathbf{R}^{n \times m}_+| \mathbf{P}\mathbf{1}\leq \mathbf{a}, \mathbf{P}^T\mathbf{1}\leq \mathbf{b}, \langle\mathbf{P}, \mathbf{1}\rangle_F = s\}$, $C$ is a cost matrix that has infinite entries, $\mathcal{M}_{ijkl}= 
(\mathbf D^{\mathrm{A}}_{ik} - \mathbf D^{\mathrm{B}}_{jl})^2$ if $|\mathbf D^{\mathrm{A}}_{ik} - \mathbf D^{\mathrm{B}}_{jl}| \leq \rho$, $=\infty$ if $|\mathbf D^{\mathrm{A}}_{ik} - \mathbf D^{\mathrm{B}}_{jl}| > \rho$.
In addition to the molecular alignment problem, the fsGW framework can also be applied to other alignment problems in chemistry. For example, the comparison of structurally similar molecules is an important problem in chemistry, especially in drug discovery. When two similar molecules have a common substructure, we can view the problem of finding the maximum common substructure as a general fsGW problem whose goal is to transport the maximal number of atoms from one molecule to another.
Looking forward, the application of optimal transport theory to other molecular alignment problems is highly promising.

\section*{Data and Code Availability}
Code and data are available at \href{https://github.com/weixiaoqimath/otmol}{https://github.com/weixiaoqimath/otmol}. \\

\section{Acknowledgements}
This work is partly supported by National Science Foundation grants DMS2151934 (Z.C.) and DMS2142500 (Y. Zhao) and U.S National Institutes of Health grants R01GM152494 (Z.C.) and R35-GM127040 (Y. Zhang). We thank NC State High Performance Computing Services Core Facility, NYU-ITS and the Simons Center for providing computational resources. X.D. acknowledges partial support from a graduate fellowship from the Simons Center for Computational Physical Chemistry (SCCPC) at NYU.


\end{document}